\documentclass[aip,rsi,reprint,amsmath,amssymb,floatfix]{revtex4-1}
\usepackage[utf8x]{inputenc}
\usepackage{siunitx}
\usepackage[breaklinks=true,colorlinks=true,linkcolor=blue,urlcolor=blue,citecolor=blue]{hyperref}
\usepackage{graphicx}
\usepackage{subfig}
\usepackage{nicefrac}
\usepackage{textcomp}
\usepackage{ragged2e}
\usepackage{booktabs}

\DeclareCaptionJustification{justifying}{\justifying}
\DeclareSIUnit\permille{\text{\textperthousand}}

\begin{document}

\def\figureautorefname{Fig.}
\def\subfigureautorefname{Fig.}
\def\equationautorefname{Eq.}
\def\tableautorefname{Tab.}

\bibliographystyle{apsrev4-1}

\title{An electrostatic in-line charge-state purification system for multicharged ions in the kiloelectronvolt energy range}

\author{Daniel Schury}
\email[]{Daniel.Schury@insp.upmc.fr}
\affiliation{Institut des Nanosciences de Paris, Sorbonne Universit\'{e}, CNRS UMR 7588, 4 place Jussieu, 75252 Paris, France}
\author{Ajit Kumar}
\affiliation{Institut des Nanosciences de Paris, Sorbonne Universit\'{e}, CNRS UMR 7588, 4 place Jussieu, 75252 Paris, France}
\author{Alain M\'{e}ry}
\affiliation{CIMAP, CEA/CNRS/ENSICAEN/Universit\'{e} de Caen Normandie, 6 Boulevard du Mar\'{e}chal Juin 14050 Caen, France}
\author{Jean-Yves Chesnel}
\affiliation{CIMAP, CEA/CNRS/ENSICAEN/Universit\'{e} de Caen Normandie, 6 Boulevard du Mar\'{e}chal Juin 14050 Caen, France}
\author{Anna L\'{e}vy}
\affiliation{Institut des Nanosciences de Paris, Sorbonne Universit\'{e}, CNRS UMR 7588, 4 place Jussieu, 75252 Paris, France}
\author{St\'{e}phane Mac\'{e}}
\affiliation{Institut des Nanosciences de Paris, Sorbonne Universit\'{e}, CNRS UMR 7588, 4 place Jussieu, 75252 Paris, France}
\author{Christophe Prigent}
\affiliation{Institut des Nanosciences de Paris, Sorbonne Universit\'{e}, CNRS UMR 7588, 4 place Jussieu, 75252 Paris, France}
\author{Jean-Marc Ramillon}
\affiliation{CIMAP, CEA/CNRS/ENSICAEN/Universit\'{e} de Caen Normandie, 6 Boulevard du Mar\'{e}chal Juin 14050 Caen, France}
\author{Jimmy Rangama}
\affiliation{CIMAP, CEA/CNRS/ENSICAEN/Universit\'{e} de Caen Normandie, 6 Boulevard du Mar\'{e}chal Juin 14050 Caen, France}
\author{Patrick Rousseau}
\affiliation{CIMAP, CEA/CNRS/ENSICAEN/Universit\'{e} de Caen Normandie, 6 Boulevard du Mar\'{e}chal Juin 14050 Caen, France}
\author{S\'{e}bastien Steydli}
\affiliation{Institut des Nanosciences de Paris, Sorbonne Universit\'{e}, CNRS UMR 7588, 4 place Jussieu, 75252 Paris, France}
\author{Martino Trassinelli}
\affiliation{Institut des Nanosciences de Paris, Sorbonne Universit\'{e}, CNRS UMR 7588, 4 place Jussieu, 75252 Paris, France}
\author{Dominique Vernhet}
\affiliation{Institut des Nanosciences de Paris, Sorbonne Universit\'{e}, CNRS UMR 7588, 4 place Jussieu, 75252 Paris, France}
\author{Emily Lamour}
\affiliation{Institut des Nanosciences de Paris, Sorbonne Universit\'{e}, CNRS UMR 7588, 4 place Jussieu, 75252 Paris, France}

\date{\today}

\begin{abstract}
The performance of a newly built omega type electrostatic analyzer designed to act as an in-line charge-state purification system for ions in the \si{\kilo\electronvolt} energy range is reported. The analyzer consists of a set of four consecutive electrostatic \SI{140}{\degree} concentric cylindrical electrodes enclosed by Matsuda electrodes. This setup was recently tested and validated using O\textsuperscript{5+}, Ar\textsuperscript{9+} and Xe\textsuperscript{20+} ion beams at an energy of \SI{14}{q\kilo\electronvolt} at the ARIBE facility. A resolving power of \num{10,5} and a transmission of \SI{100}{\percent} of the desired charge state are measured allowing a good purification of incoming ion beams with charge states up to 10+ and a fairly good purification for charge states at least up to 20+. In comparison with other in-line solutions such as Wien filter, our system has the advantage of being purely electrostatic and therefore lacking common drawbacks as for example hysteresis.
\end{abstract}


\maketitle

\section{Introduction}\label{sec:Introduction}
Charged particle beams are used in many different fields, ranging from fundamental research to applied industrial and even medical applications. Therefore, it is essential to completely control the beam using ion optical elements such as deflectors and lenses. While magneto-optical elements (such as solenoids or quadrupoles) play a dominant role for high energy beams, in the low to mid energy range electrostatic elements have the advantage of being easier to build and are additionally lacking hysteresis. Electrostatic devices have the additional advantage of separating ions by their energy per charge, needing no reconfiguration for ions with different masses, since, for a simple acceleration by an extraction voltage, the ion energy is mass independent. The most common electrostatic beam deflectors are parallel plate mirror analyzers \citep{Yarnold1949}, cylindrical mirror analyzers \citep{Sar-el1967}, radial cylindrical analyzers \citep{Hughes1929}, spherical capacitors \citep{Purcell1938} and quadrupole deflectors \citep{Zeman1977}. While these deflectors are often used to simply change the ion trajectory, for example in order to be directed onto a detector, they can also be used as energy analyzers or monochromators for particles with the same charge \citep{Harrower1955}, in which case the change of the propagation axis can be an unwanted side effect. Other ion optical elements acting as energy filters like a Wien filter \citep{Wien1898} or quadrupole ion guides \citep{Paul1953} (with the latter restricted to low energies) maintain the axis of translation but introduce again more complex elements such as magnets and high frequency voltage supplies. To preserve the simplicity of a purely electrostatic system while simultaneously keeping the axis of the incoming beam we adapt the design of four consecutive \SI{140}{\degree} cylindrical deflectors \citep{Rose1990a} with Matsuda electrodes \citep{Matsuda1961} obtaining an adjustable double focusing in-line charge state purification system. The additional Matsuda electrodes create an adjustable electric field which on the optical axis resembles that of a toroidal deflector. This design allows focusing of the beam in both transversal planes without requiring expensive toroidal electrodes.

For many experiments dealing with slow multi-charged ions interacting with matter, a good control of the charge state upstream the collision is important. This is especially true in cases in which the charge state products are measured in coincidence, for example in ion-surface/2D material experiments \citep{Gruber2018, Aumayr2011, Wilhelm2017} or ion-ion studies in the low energy domain \citep{Braeuning2005,Braeuning2006}. This purification is important in our case, as ion-ion studies normally suffer already from a low signal-to-noise ratio (SNR) in the order of \num{1e-4} due to the low ion beam densities compared to the residual gas density \citep{Meuser1996}. Since the primary ion beam intensity is normally up to ten orders of magnitudes higher than the product beam intensity, even impurities of less than \SI{1}{\percent} can noticeably decrease the SNR further and should be removed from the beam as close to the interaction point as possible. The charge state purification system presented here is developed in the context of the \emph{F}ast-\emph{I}on-\emph{S}low-\emph{I}on-\emph{C}ollision project (FISIC\citep{Lamour2013}), which aims to obtain absolute electronic cross sections for ion-ion collisions between fast (\si{\mega\electronvolt/\atomicmassunit}) and slow (\si{\kilo\electronvolt/\atomicmassunit}) ions. In this regime the ion stopping power (energy transfer) is maximum, leading to the strongest effects on material modifications, including biological materials. Until today, measurements  and reliable theoretical predictions are completely lacking in this specific collision regime.

Former similar experiments performed either slow-ion/slow-ion \citep{Meuser1996,Braeuning2005,Braeuning2006} or fast-ion/plasma-discharge collisions \citep{Fite1960, Olson1977,Cayzac2017}. To be able to carry out crossed-beam experiments at different facilities, including the S\textsuperscript{3} beamline at GANIL\citep{Dechery2015,Savajols2019} or storage rings such as CRYRING@ESR \cite{Lestinsky2016b} located at the upcoming FAIR facility, a compact and versatile setup providing a pure ion beam of highly charged low energy ions is needed. To fully enclose the high energy ion beam envelope, a beam spot size of not less than \SI{5}{\milli\metre} in height is demanded. As the two ion beams are crossed perpendicular, the demands in horizontal focusing are less strict and mainly defined by the acceptance of the ion optical elements further downstream. From our numerical simulations we expect a horizontal width of less than \SI{20}{\milli\metre} to be sufficient.

In this work we first discuss the basic principle of the purification device, describing the design and its different components. The dimensions are derived for an assumed Gaussian shaped ion beam, about \SI{2}{\centi\metre} in diameter with an emittance of approximately \num{60} $\pi$ \si{\milli\metre \, \milli\radian}. These are typical values we expect to be delivered from our yet to be assembled and characterized ECR ion source. It should be noted that in cases we want to emphasize the fact that we provide the emittance $ \epsilon $ obtained by the area of the phase space ellipse $ A = \pi \epsilon $, we give the emittance in $\pi$ \si{\milli\metre \, \milli\radian}, effectively stating the area \textit{A} of the phase space ellipse instead of the emittance itself, in accordance with \cite{Forrester1988}. Furthermore, if not stated otherwise, we provide the emittance of the phase space ellipse covering \SI{90}{\percent} of the beam ensemble. Second, we describe the experimental findings achieved during a beam time at the ARIBE facility at GANIL/Caen, which will not be the source of low energy ions we will use for the FISIC experiment mentioned before, but delivering beams suitable for first performance tests. Tests on transmission, separation power and comparisons with numerical simulations will be shown.

\section{Operational principle and design}\label{sec:operation}
Even under good vacuum conditions, especially highly-charged low-energetic ion beams undergo atomic reactions with residual gas particles. In the inelastic regime occurs mostly electron capture from the residual gas. At residual gas pressures in the order of \SI{1e-8}{\milli\bar}, transport distances of a few meters and cross sections in the order of a few \SI{1e-15}{\per\square\centi\metre} the fraction of capture products in the beam can be up to a few percent of the primary beam and lead to a source of systematic error in the experiment. For collisions with \si{\kilo\electronvolt}-beams the momentum exchange onto the ion is small\citep{Roncin1984,Barat1987} ($ \ll \SI{1}{\percent} $) and can be neglected in first approximation. As a result the ion does only change its charge state but not its kinetic energy. To clean the ion beam from these unwanted products, we adapted the design of four consecutive \SI{140}{\degree} cylindrical deflectors with Matsuda electrodes, which was initially used for electron spectrometers \citep{Rose1990a}, to separate ions with the same kinetic energy but different charge states.
\begin{figure}[t!]
	\includegraphics[width=\linewidth]{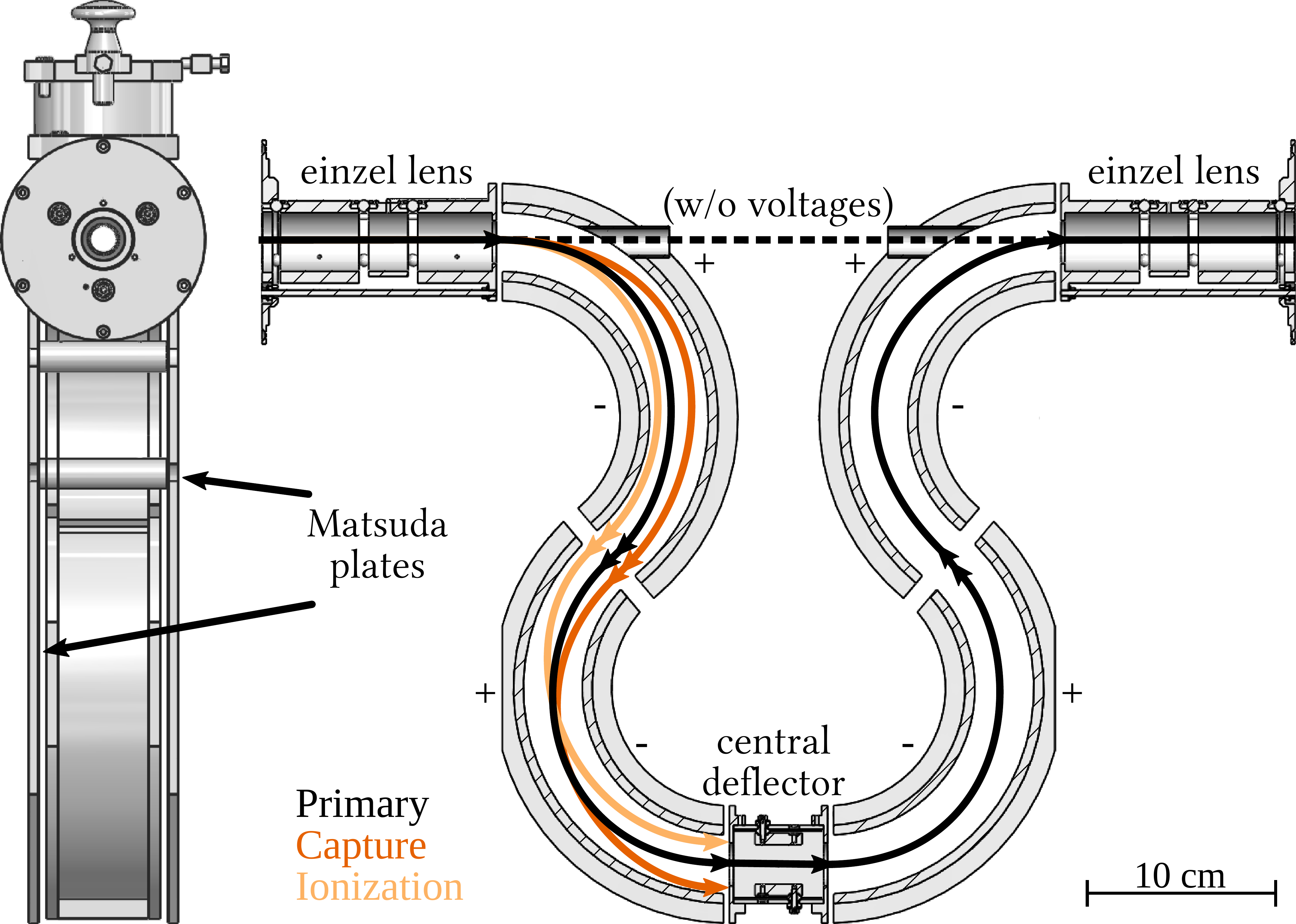}
	\caption{\label{fig:omega_scheme}Side view and schematic cut through the purification system. The cylindrical deflectors are enclosed between Matsuda electrodes. The polarity of the deflector electrodes for positive ion beams is indicated by plus and minus signs. The black beam resembles the primary ion beam which transits the structure at the effective radius. Beams with lower (capture, dark orange) or higher (ionization, light orange) charge state are sorted out by hitting the wall of the deflector.}
\end{figure}

By equating the electrostatic and centrifugal force, we can calculate the radius $ R $ of the trajectory for a particle with the kinetic energy $ E_{\mathrm{kin}} $ and the charge $ q $ which travels inside a homogeneous electric field $ F = \frac{U_0}{d} $, defined by the voltage difference $ U_0 $ and the cathode gap $ d $, as
\begin{align}\label{eq:effective_radius}
	R = \frac{2 E_{\mathrm{kin}}}{q F}
\end{align}
So as a result, ions originating from the same primary beam and hence having the same kinetic energy $ E_{\mathrm{kin}} $ but, for example due to the aforementioned electron capture, different charge states $ q $ have different radii in a cylindrical deflector with fixed $ F $. Hence we specify the energy of the ion beam in the unit of \si{q\kilo\electronvolt}, to emphasize that parasitic ions, although having the same kinetic energy will have different trajectories due to their different charge state. Based on this principle we have constructed an omega shaped analyzer to guide ions with the desired charge state while blocking the undesired ones.

In \autoref{fig:omega_scheme} a schematic cut through the described purification system is shown. We designed the purification system with an inner radius of the cylindrical deflectors of $ R_i = \SI{75}{\milli\metre} $ and an outer radius of $ R_o = \SI{105}{\milli\metre} $, resulting in an effective radius of $ R_e = \SI{90}{\milli\metre} $. Specified are the outer radius of the inner plate and the inner radius of the outer electrode, as these are the relevant values which define the electric fields. The bending angle is \SI{140}{\degree} and the electrode height \SI{58}{\milli\metre}. The omega analyzer is mounted vertically in a cylindrical vacuum chamber so that ions entering it are deflected downwards first. On the beam axis, a hole with \SI{14}{\milli\metre} diameter is included in the first and last outer deflector electrode (see \autoref{fig:omega_scheme}) so that incoming ion beams can go straight through the analyzer in case no voltages are applied on the cylindrical electrodes. For this use-case, an additional movable intermediate Faraday cup can be positioned in the beam trajectory to measure the beam intensity during initial beam alignment. It should be noted that the operation of the omega analyzer is limited to the transport of ions with kinetic energy not more than \SI{30}{q~\kilo\electronvolt} as the cylindrical deflector electrodes are connected to their high voltage power supplies with standard \num{10}-\si{\kilo\volt}-SHV-connectors. Higher energies would demand voltages higher than the \SI{10}{\kilo\volt}. The relatively large size of the analyzer is necessary in order to achieve full transmission for high emittance ion beams up to at least $ \epsilon_{\SI{90}{\percent}} = \SI{60}{\milli\metre \, \milli\radian} $. Therefore, the dimensions of the omega (aperture diameter, distances between the electrodes, lens diameters) were optimized in order to achieve the maximum transmission for large emittance ion beams.

\begin{figure}
	\includegraphics[width=.85\linewidth]{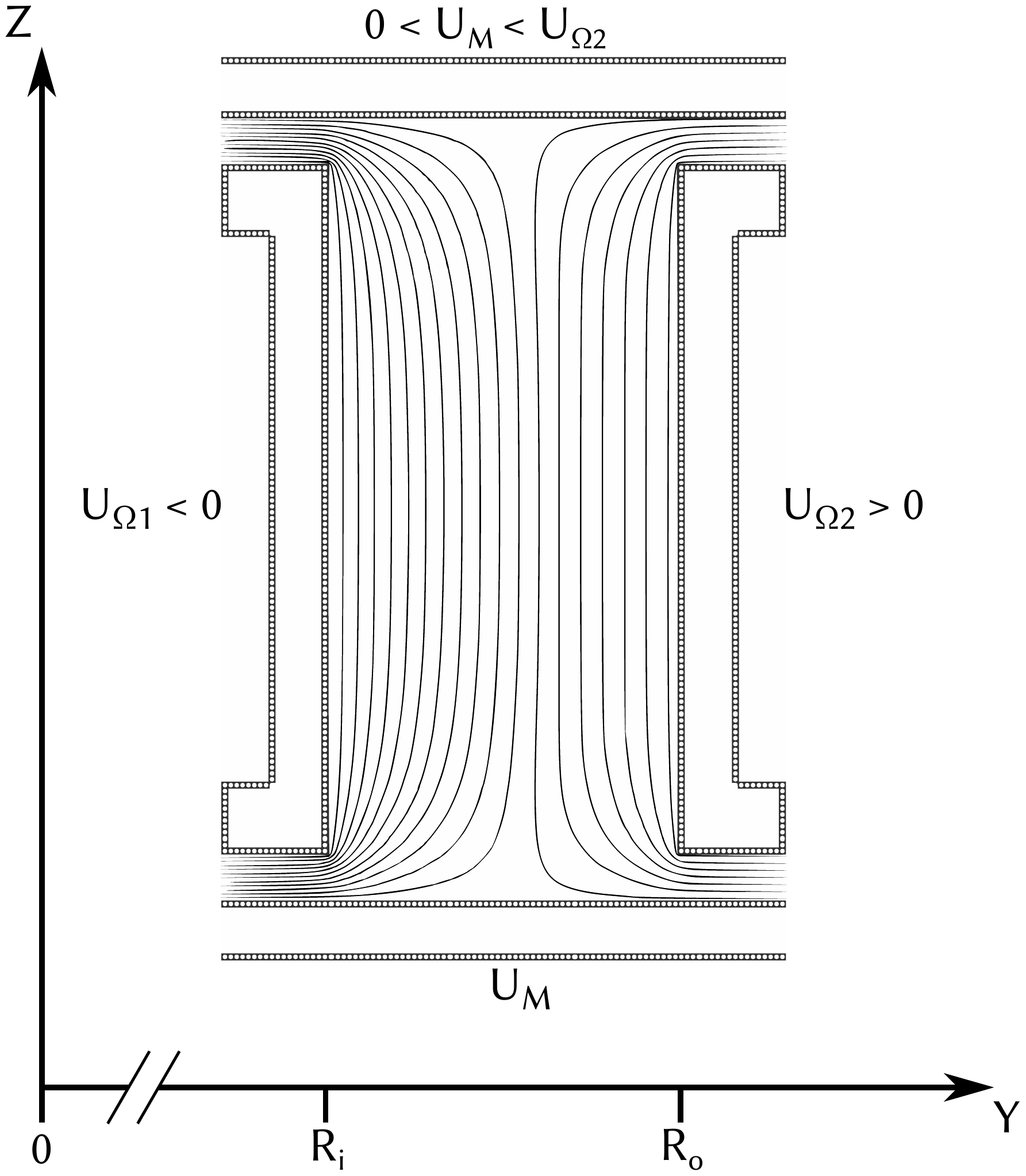}
	\caption{\label{fig:matsuda}Cut view through the analyzer, showing the inner (right) and outer (left) cylindrical electrodes as well as the two Matsuda electrodes enclosing them on top and bottom. Indicated are the equipotential lines of the electric potential which in the central region resembles that of a spherical deflector. The z-axis indicates the common axis of the cylindrical electrodes.}
\end{figure}
The use of Matsuda electrodes creates a field in the central part of the analyzer which resembles that of a toroidal deflector\citep{Matsuda1961,Matsuda1975} as illustrated in \autoref{fig:matsuda}. By changing the voltage on the electrodes, typically a few hundred volt, the toroidal factor and hence the transversal beam focusing both in the dispersive and non-dispersive plane can be adjusted\citep{Yavor1995,Yavor1998}. At the entrance and exit there is a common three-element einzel lens \citep{Aberth1974}. In the center, a pair of parallel electrodes is located between the second and third section of the analyzer. They are used to create an electrical correction field to adjust the outgoing beam direction in the vertical plane. This is especially important for higher initial charge states where according to \autoref{eq:effective_radius} the separation between the trajectories of adjacent charge states inside the analyzer is small. In this case often a correction of the beam axis is necessary. Typical voltage differences between the electrodes are not higher than \SI{3}{\kilo\volt}.

Design, optimization and analysis of the purification structure were done by carrying out ion trajectory calculations with the \emph{SIMION 3D} suite\citep{Dahl2000}. The trajectories of three ion beams with the same kinetic energy $ E_{\mathrm{kin}} $ are represented in \autoref{fig:omega_scheme} by curves of different color (black, dark orange and light orange). They only differ in their charge states: $ q_0 $ (black curve) for the ion beam labeled \emph{Primary}, $ q_0 - 1 $ (dark orange) for the ion beam labeled \emph{Capture }and $ q_0 + 1 $ (light orange) for the ion beam labeled \emph{Ionization}. In accordance with \autoref{eq:effective_radius} the electric field $ F $ is set so that the primary ion beam travels through the setup on the designed effective radius $ R_e $. The trajectory radii of the two other beams (capture and ionization) do not allow these beams to reach the exit of the setup due to the greater (capture) or smaller (ionization) radius.

From simulations we expect the analyzing power of the omega analyzer to be high enough to separate broad ($ \mathrm{FWHM} = \SI{7}{\milli\metre} $) beams with an emittance of $ \epsilon_{\SI{90}{\percent}} = \SI{60}{\milli\metre \, \milli\radian} $ for charge states up to $ q_0 \approx 30 $. For narrower ($ \mathrm{FWHM} = \SI{2,5}{\milli\metre} $) beams with an emittance of $ \epsilon_{\SI{90}{\percent}} = \SI{10}{\milli\metre \, \milli\radian} $ we expect to be able to separate charge states up to bare xenon ($ q_0 = 54 $).
\begin{figure}
	\centering
	\subfloat{%
		\includegraphics[width=0.4\textwidth]{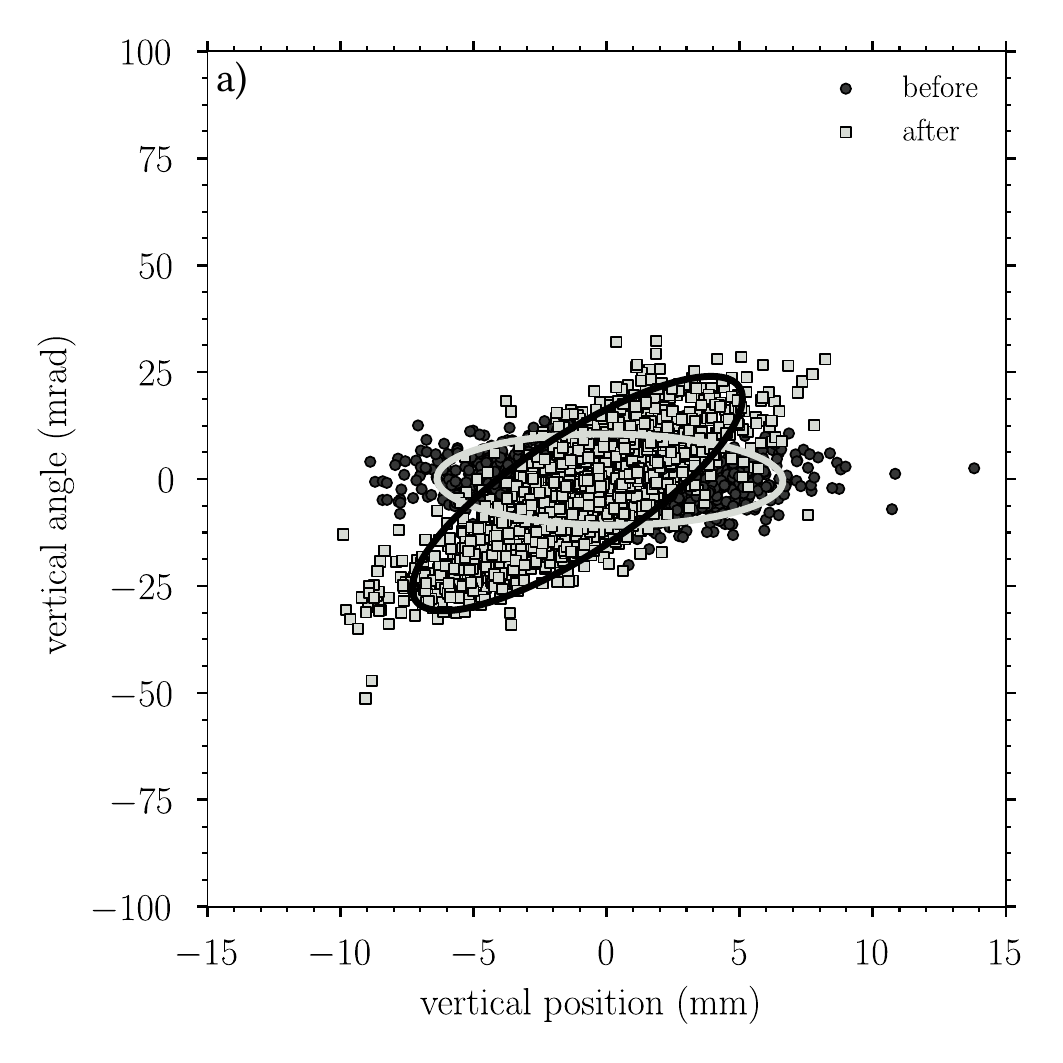}\label{fig:vert_emit}
	}\\
	\subfloat{%
		\includegraphics[width=0.4\textwidth]{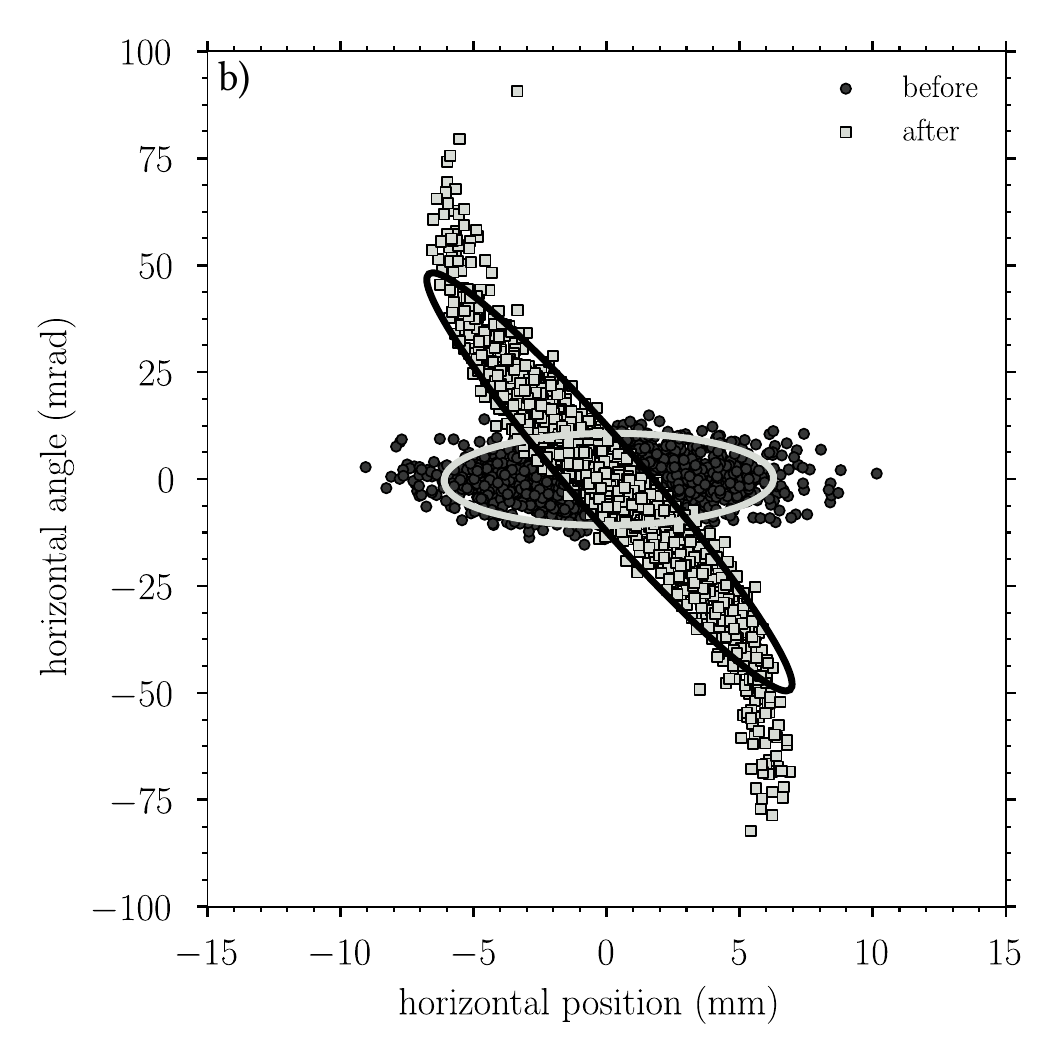}\label{fig:hor_emit}
	}
	\caption{\label{fig:before_after}a) Vertical and b) horizontal phase spaces of a simulated ion beam at the entrance of the first einzel lens (black dots) and the end of the second einzel lens (grey squares) of the omega analyzer. Simulated is a centered starting Gaussian Ar\textsuperscript{9+} ion beam with $ E_0 = \SI{14}{q\kilo\electronvolt} $, $ \Delta E = \SI{0,5}{\percent} $, $ \mathrm{FWHM} = \SI{7}{\milli\metre} $ and $ \epsilon_{\SI{90}{\percent}} = \SI{60}{\milli\metre \, \milli\radian} $. The analyzer is tuned in such a way that the beam leaves the analyzer again centered with approximately the same Gaussian distribution. As a guide to the eye, the \SI{90}{\percent} emittance ellipse is drawn.}
\end{figure}

In \autoref{fig:before_after} the simulated beam emittances of a \SI{14}{q\kilo\electronvolt} centered starting Gaussian Ar\textsuperscript{9+} ion beam with initial parameters at entering the first einzel lens of \SI{0,5}{\percent} energy dispersion and a horizontal and vertical full width at half maximum of \SI{7}{\milli\metre} and an emittance of \num{60} $\pi$ \si{\milli\metre \, \milli\radian} are shown. The analyzer is tuned in such a way that the beam leaves the analyzer again centered with approximately the same Gaussian distribution. The beam parameter are measured at the beginning of the Einzel lens before and at the end of the Einzel lens after the purification system. A parallel incoming beam leaves the analyzer divergent in the vertical and convergent in the horizontal plane while keeping its size and shape. With different sets of voltages applied on the omega purification system, we are able to control the shape and the horizontal and vertical focusing of the desired beam at an interaction region located $ \approx \SI{50}{\centi\metre} $ after the omega structure.

\section{Implementation and test}\label{sec:test}
In the following tripartite section we describe a first experimental setup of the analyzer, its results and compare the outcome with our numerical simulations. In \autoref{subsec:properties} we describe the experimental setup at the ARIBE facility at GANIL/Caen and give details about the beams we used, which characteristics (emittance and size) are close to the one expected from the FISIC project and thus fit with the technical limitations of our setup. In \autoref{subsec:scans} we give details on how one can use the omega analyzer to conduct energy-charge-scans for extended information on the beam, especially regarding the separation power of the device. Finally, we present in \autoref{subsec:rates} a comparison of the measured electron capture rates and the rates we expect from known charge exchange cross sections. Additionally we compare the measured and simulated energy-charge-scans.
\subsection{Beam properties}\label{subsec:properties}
\begin{figure}[b]
	\includegraphics[width=\linewidth]{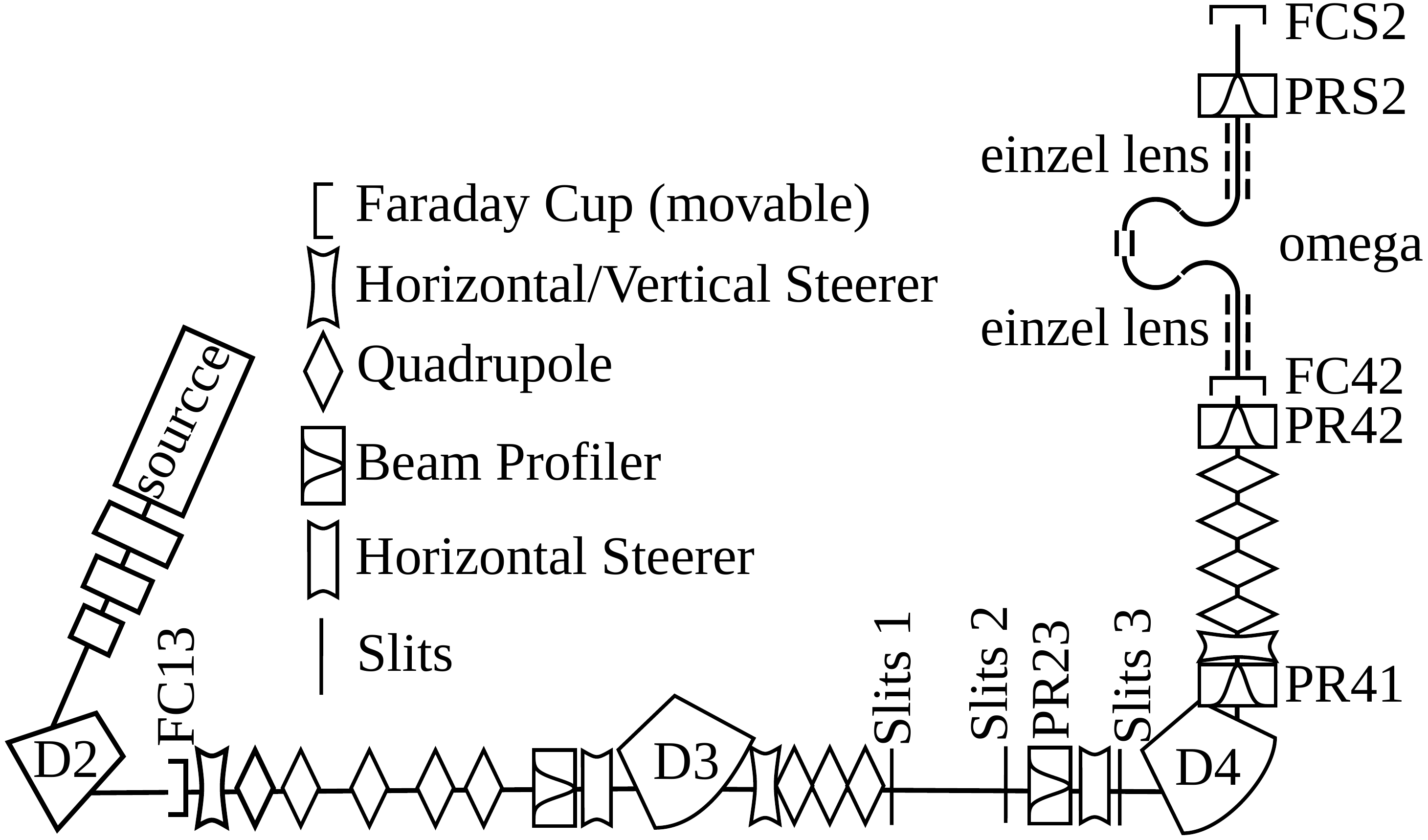}
	\caption{\label{fig:ARIBE}Schematic overview of the ARIBE beamline where the experiment took place. The elements are described in the embedded legend. Important elements are labeled accordingly.}
\end{figure}
Tests of the transmission and charge state separation abilities have been carried out at the ARIBE facility at GANIL \citep{Bernigaud2008,Rousseau2019} with three different primary ion beams: O\textsuperscript{5+}, Ar\textsuperscript{9+} and Xe\textsuperscript{20+} at \SI{14}{q\kilo\electronvolt} resulting in total energies of \SI{70}{\kilo\electronvolt}, \SI{126}{\kilo\electronvolt} and \SI{280}{\kilo\electronvolt} with typical maximum intensities of \SI{25}{\micro\ampere} for O\textsuperscript{5+} and Ar\textsuperscript{9+} and \SI{7}{\micro\ampere} for Xe\textsuperscript{20+}. The ion beams were extracted from a \SI{14,5}{\giga\hertz} ion source and charge-to-mass separated by a \SI{60}{\degree} dipole magnet. The residual gas pressure was about \SI{4e-9}{\milli\bar} without and \SI{6e-8}{\milli\bar} with active primary beam.

\begin{figure}[t]
	\centering
	\subfloat{%
		\includegraphics[width=0.5\textwidth]{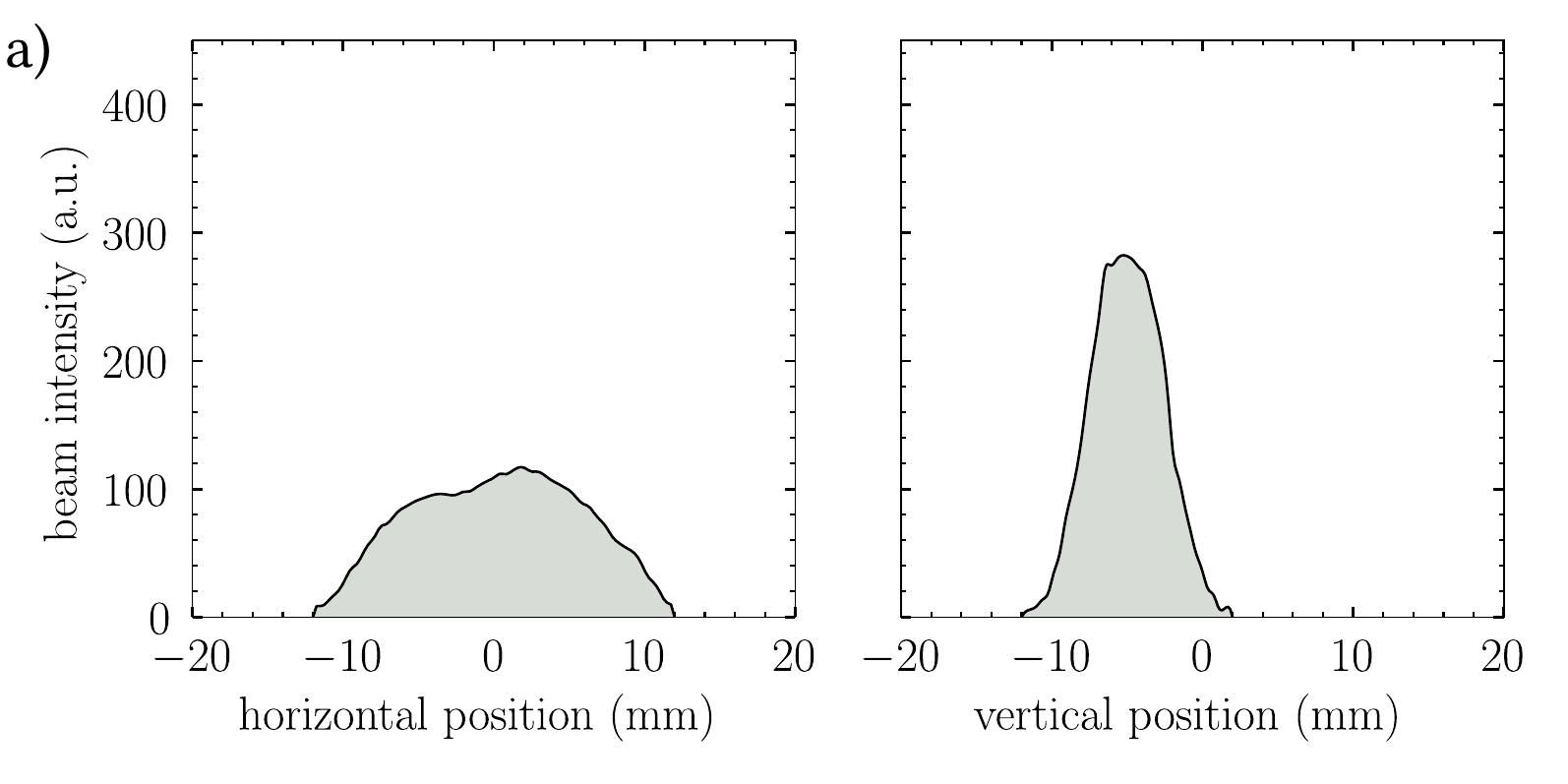}\label{fig:PR41_1}
	}\\
	\subfloat{%
		\includegraphics[width=0.5\textwidth]{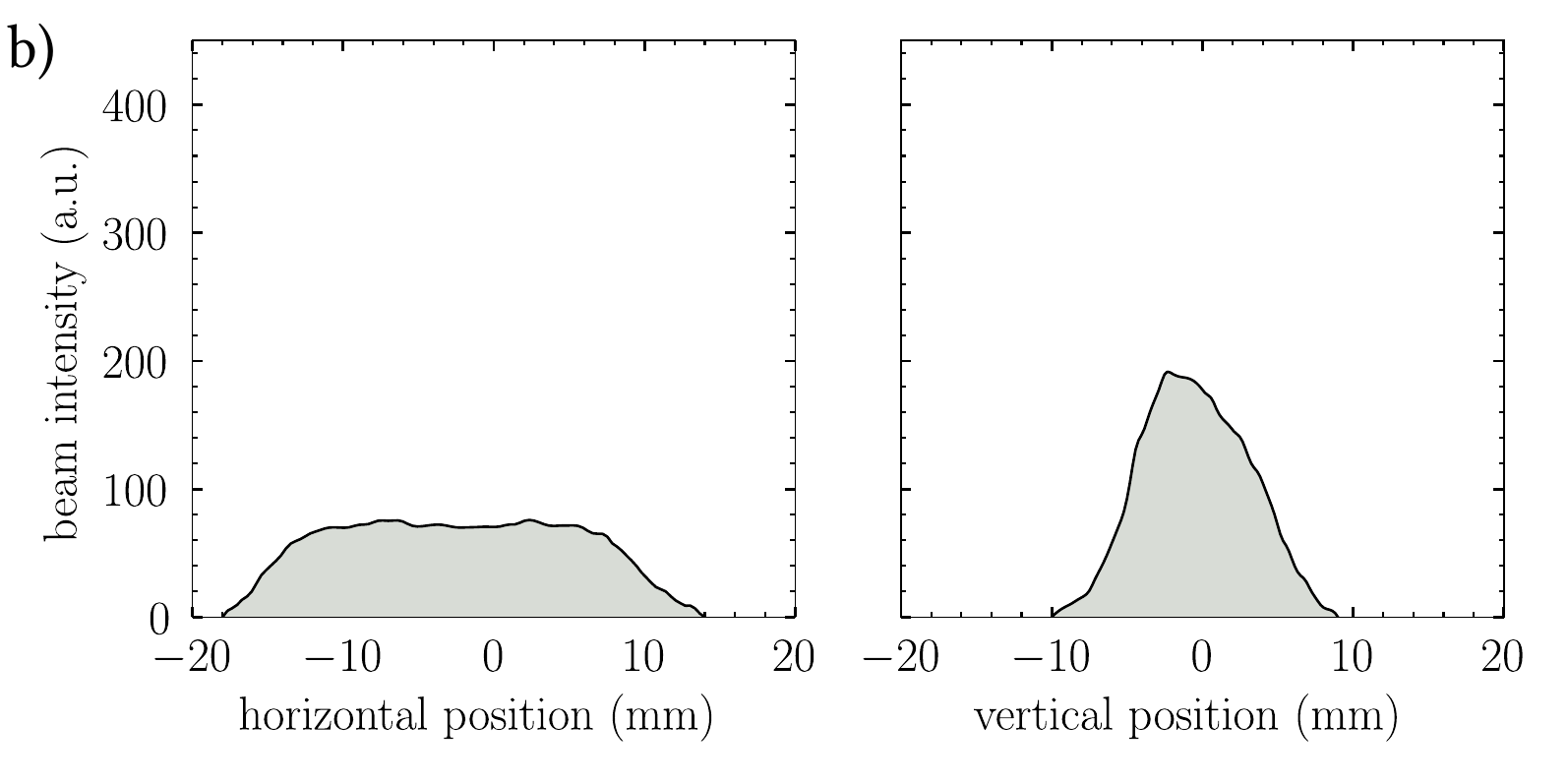}\label{fig:PR42_1}
	}\\
	\subfloat{%
		\includegraphics[width=0.5\textwidth]{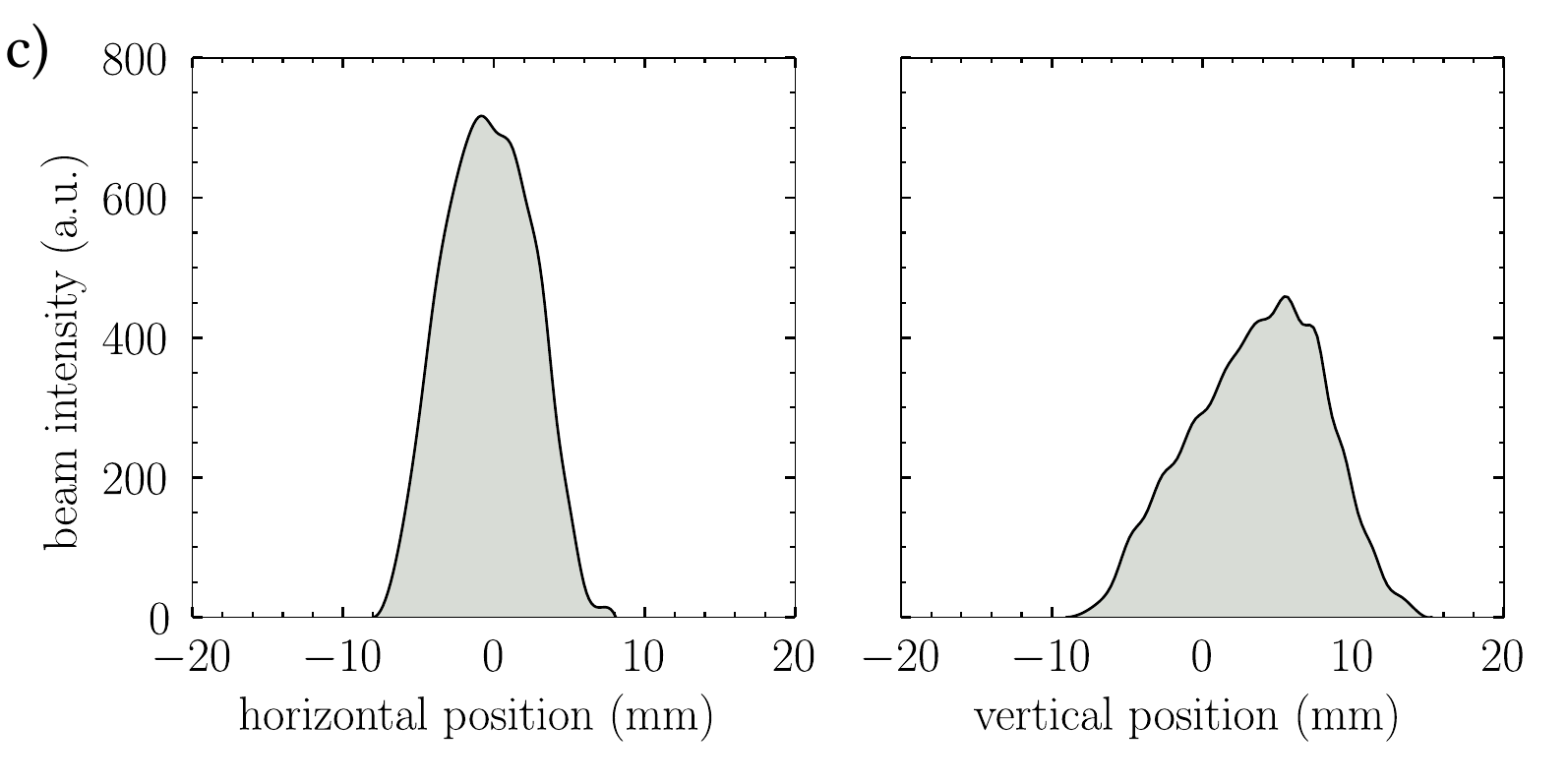}\label{fig:PRS2_1}
	}
	\caption{\label{fig:BeamProfiles}Typical \SI{126}{\kilo\electronvolt} Ar\textsuperscript{9+} beam profiles upstream (a) PR41, FWHMH = \SI{37,5}{\milli\metre}, FWHMV = \SI{9,5}{\milli\metre}; b) PR42, FWHMH = \SI{54}{\milli\metre}, FWHMV = \SI{16,5}{\milli\metre}) and downstream the omega (c) PRS2, FWHMH = \SI{3,5}{\milli\metre}, FWHMV = \SI{8,5}{\milli\metre}). The voltages on the omega were set to have \SI{100}{\percent} transmission. The slits before D4 were completely open.}
\end{figure}

\begin{table}
	\caption{\label{tab:voltages}Applied average voltage differences and minimal/maximal variation between different settings on the omega electrodes, the Matsuda electrodes and the central vertical deflector for which optimal transmission was achieved are given for the three ion beams used. In case of the omega electrodes the expected value according to \autoref{eq:effective_radius} and the values given in \autoref{sec:test} is \SI{9,333}{\kilo\volt}.}
	\begin{ruledtabular}
		\begin{tabular}{cccc}
			ion & $\Delta U_{\mathrm{omega}}\,\left(\si{\kilo\volt}\right)$  & $U_{\mathrm{matsuda}}\,\left(\si{\volt}\right)$ & $\Delta U_{\mathrm{deflector}}\,\left(\si{\kilo\volt}\right)$ \\
			O\textsuperscript{5+} & \num[separate-uncertainty = true]{9,247\pm0,108} & \num{650} & \num[separate-uncertainty = true]{2,74\pm0,058} \\
			Ar\textsuperscript{9+} & \num[separate-uncertainty = true]{9,197\pm0,203} & \num[separate-uncertainty = true]{683\pm106} & \num[separate-uncertainty = true]{1,782\pm0,776} \\
			Xe\textsuperscript{20+} & \num[separate-uncertainty = true]{9,210\pm0,136} & \num{650} & \num[separate-uncertainty = true]{2,770\pm0,001} \\
		\end{tabular}
	\end{ruledtabular}
\end{table}
The schematic setup of the ARIBE facility and the location of the omega analyzer at the end of the beamline are shown in \autoref{fig:ARIBE}. Magnetic quadrupoles and steerers allow guiding the ion beam while the intensity and profile are monitored by Faraday cups (FC) and \num{47x47} wires multi-wire beam profilers (PR) which measure horizontal and vertical beam profiles. Hence, the position, size and intensity are measured upstream the omega (PR23, PR41, PR42, FC42) and downstream (PRS2, FCS2). Typical profiles for an argon ion beam are shown in \autoref{fig:BeamProfiles} upstream (\autoref{fig:PR41_1} and \autoref{fig:PR42_1}) and downstream (\autoref{fig:PRS2_1}) the omega. In fact, by using the einzel lens just after the omega, it is possible to focus the beam to a narrow ($ \leq $\SI{1x1}{\centi\metre}) spot on the last profiler (PRS2).

\begin{figure}[t]
	\subfloat{%
		\includegraphics[width=0.5\textwidth]{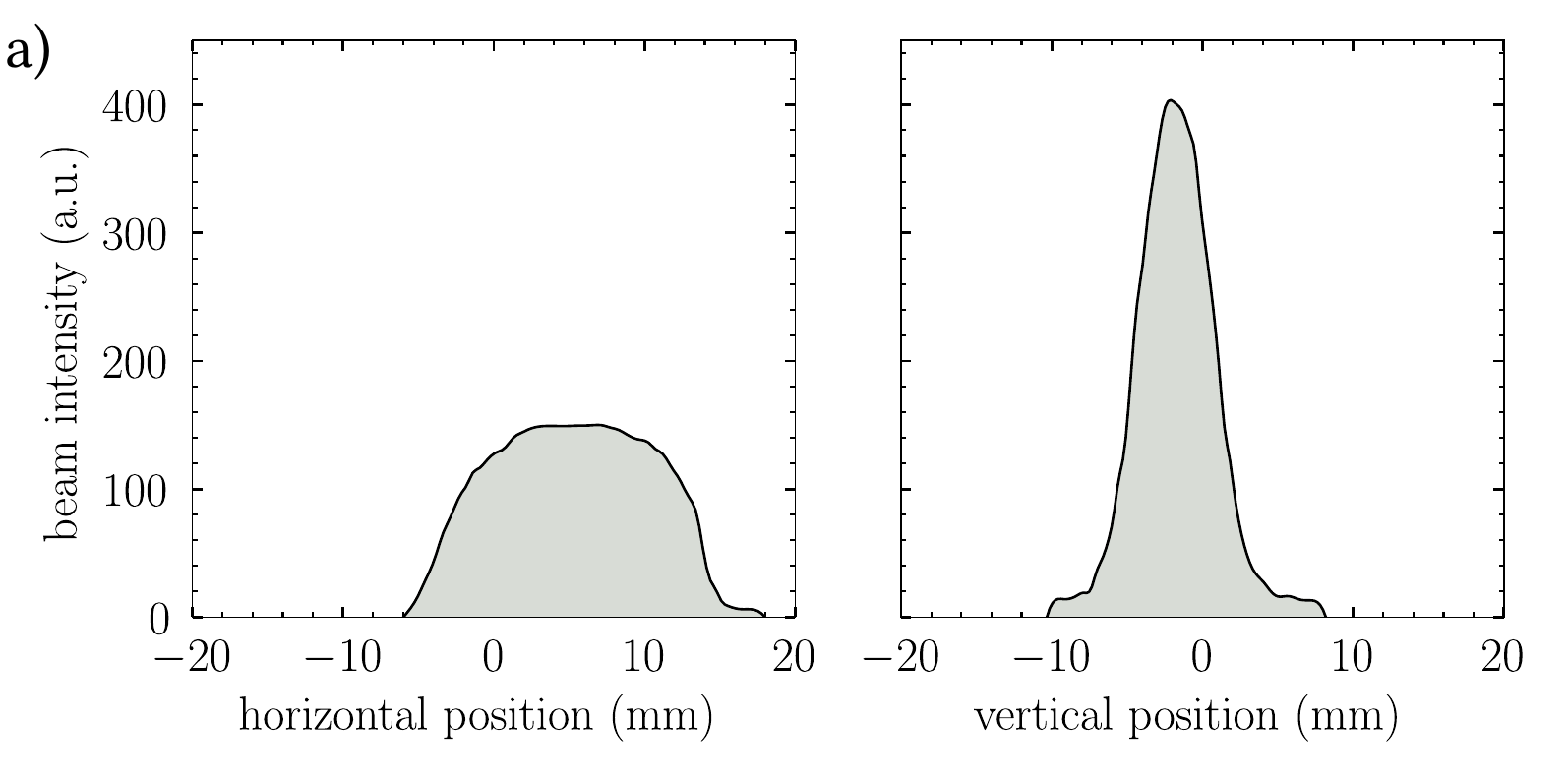}\label{fig:PR42_2}
	}\\
	\subfloat{%
		\includegraphics[width=0.5\textwidth]{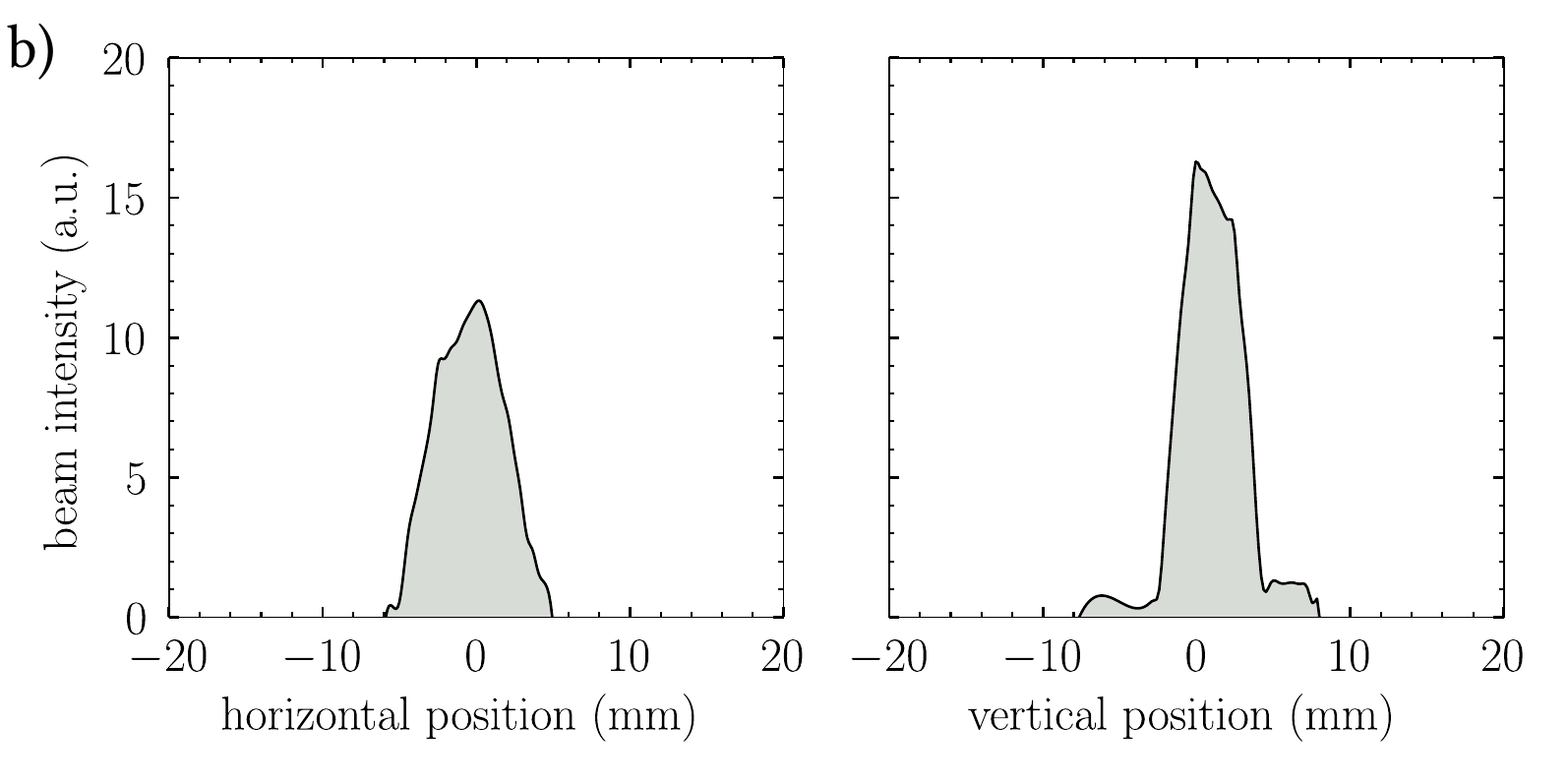}\label{fig:PR42_3}
	}
	\caption{\label{fig:BeamProfiles2}Typical \SI{280}{\kilo\electronvolt} Xe\textsuperscript{20+} beam profiles upstream of the omega (PR42), with a) fully open slits (FWHMH = \SI{33}{\milli\metre}, FWHMV = \SI{19}{\milli\metre}) and b) \SI{1x1}{\milli\metre} aperture (FWHMH = \SI{9,5}{\milli\metre}, FWHMV = \SI{7}{\milli\metre}) which cut the emittance by roughly a factor of 10.}
\end{figure}

Additionally, a series of three adjustable rectangular slits (spaced \SI{90}{\centi\meter} apart from each other) located before the last magnetic dipole (D4) can be used to reduce the emittance of the beam and in parallel its intensity and size. This possibility was used only in the case of the xenon ion beam. In \autoref{fig:BeamProfiles2} beam profiles of a xenon beam are shown. In \autoref{fig:PR42_2} the slits are fully open, in \autoref{fig:PR42_3} slits 1 and 3 are closed to \SI{1x1}{\milli\metre}. After closing slits 1 and 3, slit 2 is closed only to the point that no noticeable intensity drop is visible. This procedure gets rid of the farthest outliers of the beam which contribute most to a high beam emittance. In consideration of the beam size change after closing the slit apertures (\autoref{fig:BeamProfiles2}) we estimate a reduction of the emittance of roughly a factor of ten, while the ion beam intensity drops down to \SI{2}{\percent} of the initial value. This reduction in emittance and intensity is observed upstream of the omega analyzer, on the profiler PR42.

It should be noted that a transmission of the omega analyzer of \SI{100}{\percent} within the experimental uncertainty integrated over all charge states and especially for the primary charge state is reached in any case. The applied voltage differences on the omega electrodes as given in \autoref{tab:voltages} deviate less than \SI{1,5}{\percent} from the expected value, also the tendency goes toward slightly lower values than calculated. This deviation may be caused by alignment errors, construction accuracy, fringe field effects or uncertainty of the ion beam energy. Furthermore one has to take into account that in a cylindrical deflector the potential on the central trajectory is not zero for symmetric voltage values, effectively slowing down the ions on entering and therefore requiring a slightly lower field as predicted by \autoref{eq:effective_radius}. The applied voltage difference was quite symmetrically distributed between the inner and outer electrode. The average difference between the absolute voltage on the electrodes was \SI{4,4}{\percent}, the maximum difference was \SI{7,17}{\percent}. For oxygen and xenon a medium voltage difference of around \SI{2,7}{\kilo\volt} was applied on the central vertical deflectors, whereas for argon a slightly lower value of \SI{1,8}{\kilo\volt} was used to compensate and correct the trajectory, which in exchange deviated more strongly between different scans. The voltage on the Matsuda electrodes was \SI{650}{\volt} for the oxygen and xenon beam. For the argon beam the applied voltage was \SI[separate-uncertainty = true]{683\pm106}{\volt}.
\begin{figure}
	\subfloat{
		\includegraphics[width=0.9\linewidth]{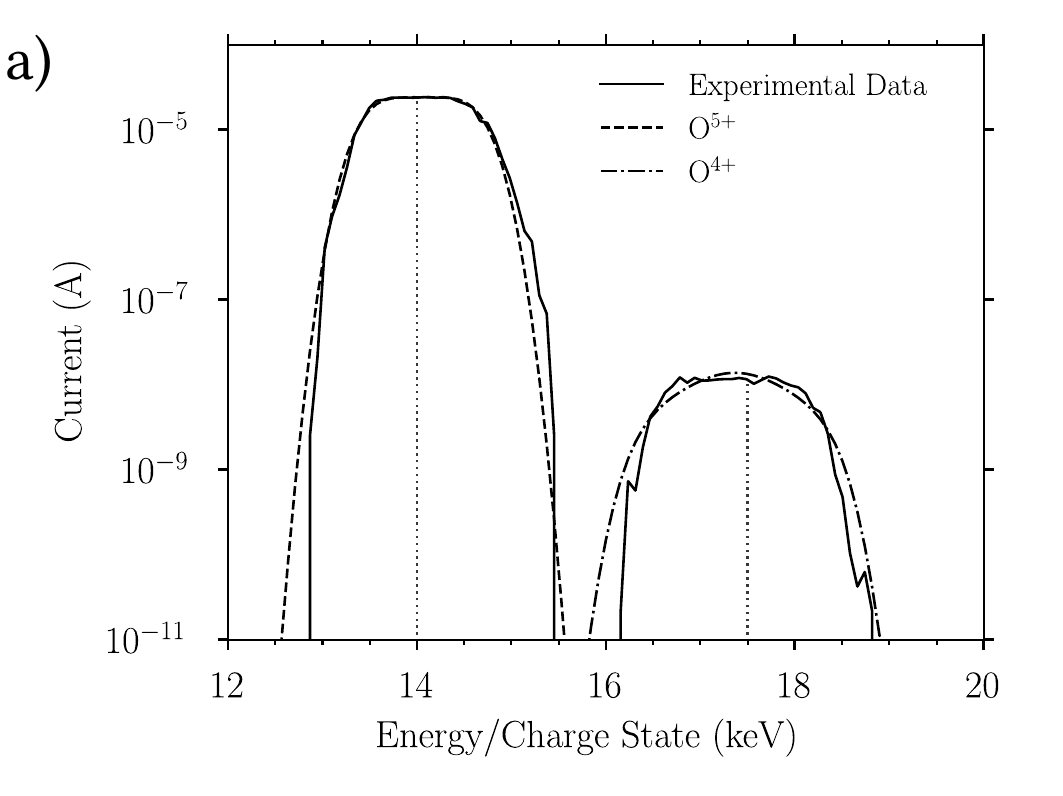}
	}\\
	\subfloat{
		\includegraphics[width=0.9\linewidth]{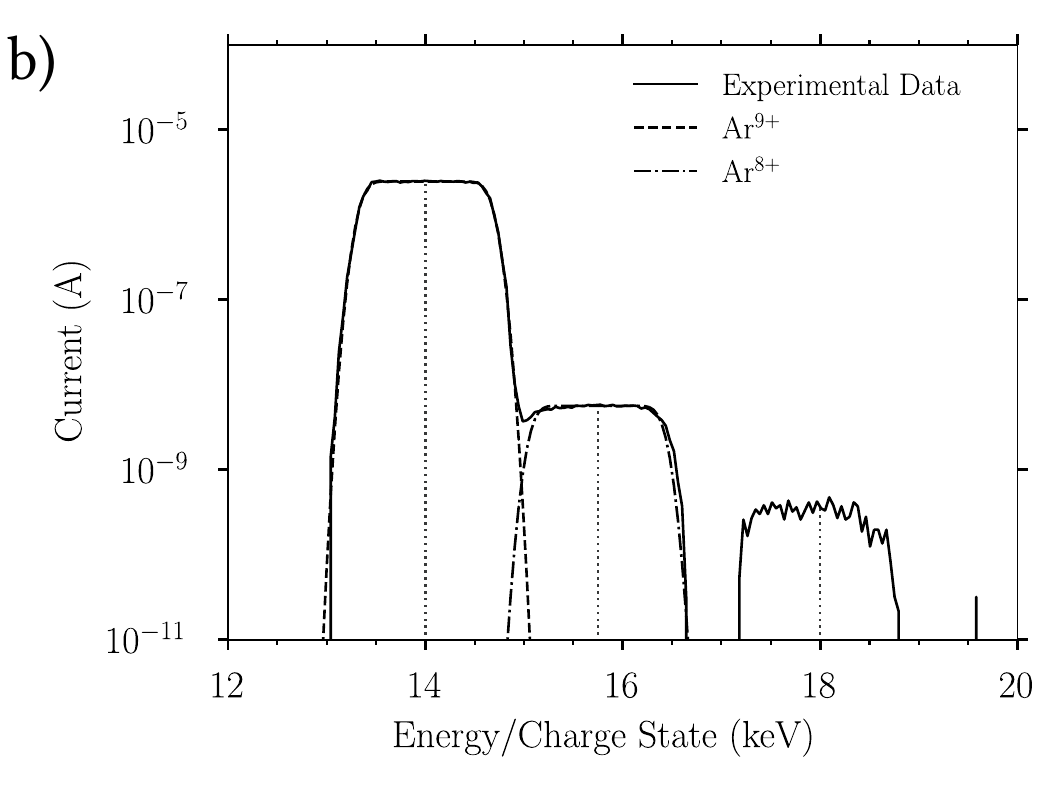}
	}
	\caption{Energy scans of a) O\textsuperscript{5+} and b) Ar\textsuperscript{9+} primary beams (solid black line). The dashed line is a fit of the primary charge state $ q_0 $, the dot-dashed line of the single electron capture charge state $ q_0 - 1 $. Vertical dotted lines represent the expected center of each charge state. The energy scale is derived by \autoref{eq:energy}.\label{fig:omega_scan}}
\end{figure}
\subsection{Scans and resolving power}\label{subsec:scans}
\begin{figure}
	\subfloat[\label{fig:xenon1}]{%
		\includegraphics[width=0.9\linewidth]{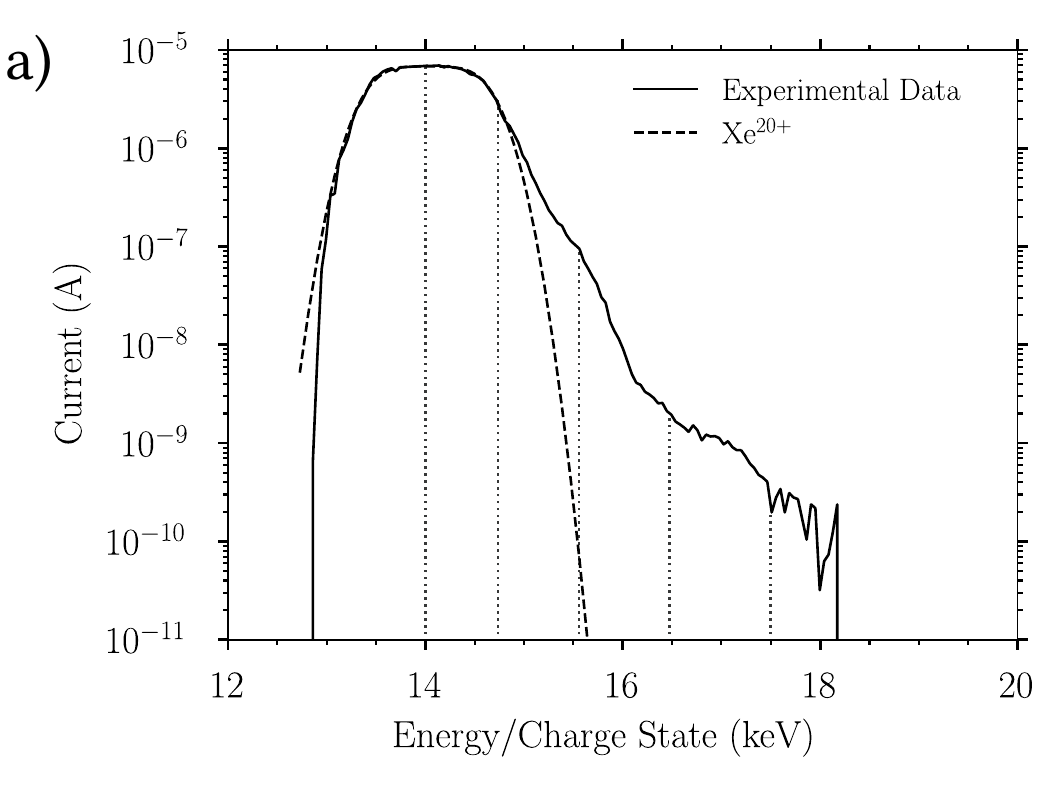}
	}\\
	\subfloat[\label{fig:xenon2}]{%
		\includegraphics[width=0.9\linewidth]{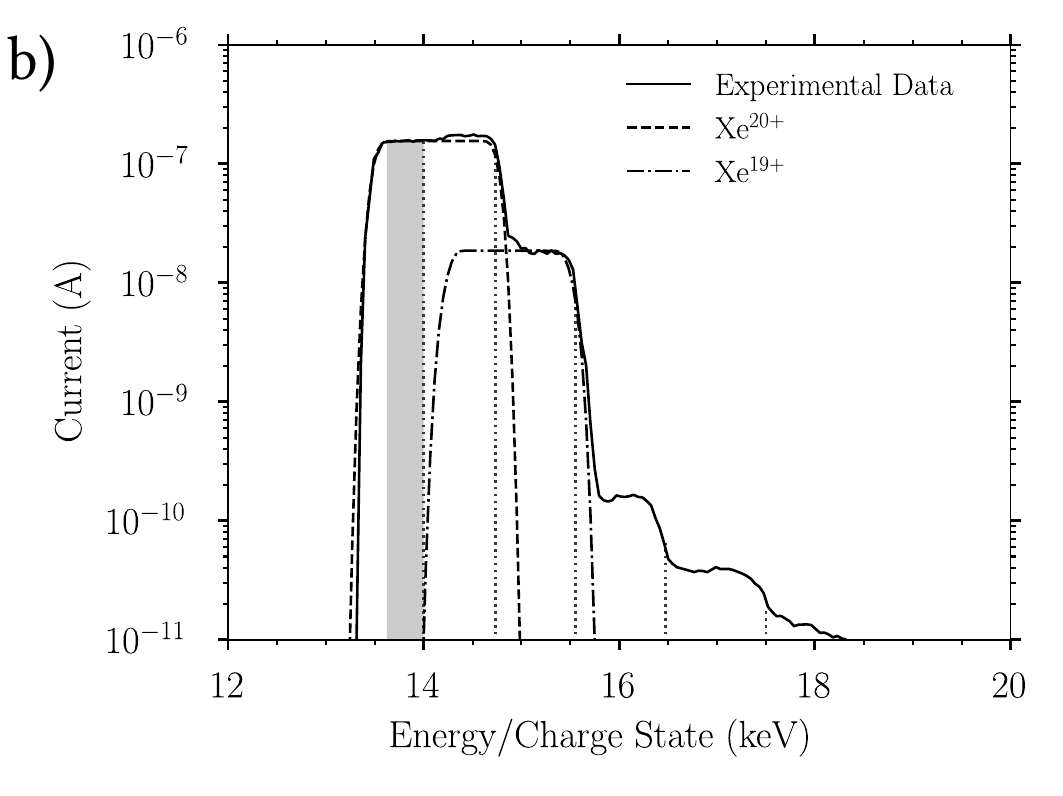}
	}
	\caption{\label{fig:Xe_broad}Energy scans of the Xe\textsuperscript{20+} beam a) before and b) after the reduction of the emittance by about a factor of ten, line scheme as in \autoref{fig:omega_scan}. The gray rectangle marks the energy range of the scan for which no overlapping of the peaks occurs.\label{fig:omega_scanXe}}
\end{figure}

In the following we introduce an energy-scan procedure, which allows us to experimentally determine the resolving power and identify the presence of unwanted charge states in the incoming primary beam. First, the voltages of the purifier are tuned to get optimal transmission of \SI{100}{\percent} within the experimental uncertainty. These values are used as initial values $ E_0 = \SI{14}{\kilo\electronvolt} $ with $ q_0 \in \{5,9,20\} $. Second, the transmitted energy and accordingly the voltages $ U^i $ are scaled as
\begin{align}
	E_q &= \frac{E_0 q_0}{q}\label{eq:energy}\\ 
	U^i_q &= \frac{U^i_0 q_0}{q}\label{eq:voltage}
\end{align}
Hereby $ U^i_0 $ represents the voltage on either the cylindrical deflector electrodes, the Matsuda electrodes or the central deflector electrodes for which optimal transmission of a beam with energy $ E_0 $ and charge state $ q_0 $ is achieved. $ U^i_q $ hence corresponds to the voltage, for which a beam with the same energy $ E_0 $ but charge state $ q $ has optimal transmission in the analyzer. To achieve scanning, at least the voltages on the cylindrical deflector electrodes have to be set following this scheme. We plot the measured current in the Faraday cup after the analyzer for every voltage step as a function of the transmission energy $ E_q $ as calculated by \autoref{eq:energy}. Then we remove a possible offset by shifting it to center the first peak at \SI{14}{\kilo\electronvolt} to obtain scans as shown in \autoref{fig:omega_scan} to \autoref{fig:sim_res}. The dotted vertical lines mark the expected center for each observed charge state. For each point in the scans the voltages on the cylindrical electrodes, the Matsuda electrodes and the central deflector electrodes are set according to \autoref{eq:voltage}.

The scans result in broad peaks with a plateau-like maximum, where the intensity stays approximately constant over a certain voltage range. The plateau of the scan arises partly from the size of the Faraday cup (FCS2), which with \SI{35}{\milli\metre} diameter is larger than the focused ion beam, partly from the design of the analyzer, which was built to have a broad acceptance range. The primary ($ q_0 $) and first capture ($ q_0-1 $) peaks are fitted with the sum of two Gaussian normal distributions each convoluted with a rectangular function to take into account the large acceptance.

We calculate the resolution of the analyzer similar as in mass spectrometry\citep{Murray2013} as $ \nicefrac{E}{\Delta E} $, where $ \Delta E $ is the full width at half maximum (FWHM) of the measured peak. By this method we obtain an energy resolution of approximately \num{10,5} for all three ions, which is in accordance with the resolution we expect from our simulations as shown in \autoref{fig:sim_res}. This means, following the canonical interpretation of the resolution, $ q_0 = 10 $ would be the highest initial charge state for which a clear separation could be achieved. However, separation can still be achieved for such high charge states with $ q \geq \num{10} $ due to the fact that the peaks in the energy scan are not Gaussian but rectangular shaped with a sharp cutoff. This can be clearly seen in the case of the xenon beam (\autoref{fig:Xe_broad}) where the neighboring charge states overlap in about half of their width. Since the omega purification system has a broad acceptance range of voltages for which a beam is transported through the structure it is possible to transport two neighboring charge states in parallel. So it becomes necessary to carefully adjust and reassure the voltages with the help of the energy scans to avoid charge state contamination. By carefully adjusting the voltages towards the edge of the scan we are able to use the outer part of the peak, which even for Xe\textsuperscript{20+} is background free while still at almost full intensity. This area is marked in gray in \autoref{fig:xenon2}. It seems safe to state that the real resolving power is at least a factor of two higher by detuning the electrode voltages towards lower energies, even though due to the unfortunate shape of the scans it is difficult to derive a consistent analytical number representing the higher resolution of the device. Nevertheless, the resolution is high enough to reduce the impurities in the primary beam after the omega analyzer below the mentioned required ten orders of magnitude, as can be deducted from \autoref{fig:omega_scan} and \autoref{fig:omega_scanXe}.

The importance of reducing the emittance for higher charge states is demonstrated in comparison between the energy scans for the xenon beam in \autoref{fig:Xe_broad}, which were carried out with slits either fully open (\autoref{fig:xenon1}) or closed down to \SI{1x1}{\milli\metre} (\autoref{fig:xenon2}). While the different charge states are initially hard to distinguish, after closing the slits they become clearly differentiated. By inserting the slits, we estimate the emittance after the slits to be reduced by roughly a factor of ten while the intensity dropped to \SI{2}{\percent}.

\subsection{Capture rates and simulations}\label{subsec:rates}
\begin{table}[b]
	\caption{\label{tab:proportion}Overview of the measured proportions of primary charge state and charge states originating from electron capture in the residual gas at a background pressure of \SI{6e-8}{\milli\bar}.}
	\begin{ruledtabular}
		\begin{tabular}{cc|cc|cc}
			\multicolumn{2}{c|}{Oxygen} & \multicolumn{2}{c|}{Argon} & \multicolumn{2}{c}{Xenon} \\
			$ q $ & fraction & $ q $ & fraction & $ q $ & fraction \\
			5 & $ \sim $\num{1,00e+0} & 9 & \num{9,97e-1} & 20 & \num{8,95e-1} \\
			4 & \num{4,82e-4} & 8 & \num{2,52e-3} & 19 & \num{1,05e-1} \\
			&  & 7 & \num{1,81e-4} & 18 & \num{9,16e-4} \\
			&  & 6 & \num{4,21e-5} & 17 & \num{2,32e-4} \\
			&  &  &  & 16 & \num{6,03e-5} \\
		\end{tabular}
	\end{ruledtabular}
\end{table}
We were able to deduce the electron capture probability due to collisions with residual gas along the ARIBE beamline by taking scans such as in \autoref{fig:omega_scan} and \autoref{fig:omega_scanXe} and by comparing the relative heights of the peaks of the different charge states. As expected for such low ion beam energies and residual gas densities, no further ionization is visible within the experimental sensitivity. The results are presented in \autoref{tab:proportion}. All product ions have a portion of the primary beam of at most \SI{2,5}{\permille}, with Xe\textsuperscript{19+} being the only exception with a portion of about \SI{10}{\percent}.

To validate the results, we compare them with the expected ratios for single capture utilizing literature values for the charge exchange cross section (CEC)\citep{Phaneuf1987, Crandall1980}. Since the residual gas composition is not known, we use a simplified model composed of \SI{100}{\percent} H\textsubscript{2}, which should account for more than \SI{95}{\percent} of the residual gas\citep{Huber1990} composition. By this choice the capture rate will be underestimated, as we ignore the presence of heavier gas species from whom electron capture has higher cross sections. We calculate the ratio $ r $ as
\begin{align}
	r &= 1-e^{-\sigma \rho l}\\
	\rho &= \frac{p N_\mathrm{A}}{R T}
\end{align}
with the CEC $ \sigma $, the length of the interaction volume $ l = \SI{3,5}{\metre} $, the vacuum pressure $ p = \SI{6e-8}{\milli\bar} $, the Avogadro constant $ N_{\mathrm{A}} $, the gas constant $ R $ and the temperature $ T = \SI{300}{\kelvin} $. The obtained results are given in \autoref{tab:proportion2}.

For oxygen and argon the calculated electron capture rates are within a factor of two of the experimental findings, a good agreement for such a rough estimation. The cross section for Xe\textsuperscript{20+} is unknown so that we had to extrapolate it from known cross sections of lower charge states up to $ q=\num{12} $. By this method we estimate a ratio about an order of magnitude lower than measured. This factor of ten is unlikely to be caused by the uncertainties of our estimation, especially since for oxygen and argon we overestimate the ratio, not underestimate it. Most likely the second dipole magnet D4 had an insufficient separation power to filter out the capture ions of the higher charge states before the omega analyzer. By failing so the actual capture distance would be approximately three times larger, shifting the result into the error range of our estimation. Another effect which could have amplified this finding could be an increased charge exchange at the slit surfaces before D4, even though one would expect mainly multiple capture in this case. Still we can not satisfactorily explain the huge capture probability for the xenon beam.
\begin{table}
	\caption{Comparison of the measured and expected ratio of single capture products in the beams. Unit of the charge exchange cross section (CEC) is \si{10^{-16}\,\centi\metre\squared}.}\label{tab:proportion2}
	\begin{ruledtabular}
		\begin{tabular}{ccccc}
			\toprule 
			ion & $ \sigma_{\mathrm{CEC}} $ & $ r_{\mathrm{theo}} $ & $ r_{\mathrm{exp}} $ & $ \nicefrac{r_{\mathrm{exp}}}{r_{\mathrm{theo}}} $ \\ 
			O\textsuperscript{5+}& \num{32.5} & \num{1.1e-3} & \num{4,8e-4} & \num{0,43} \\ 
			Ar\textsuperscript{9+}& \num{56.3} & \num{2.8e-3} & \num{2,5e-3} & \num{0,89} \\ 
			Xe\textsuperscript{20+}& \num{217,5} & \num{1,1e-2} & \num{1,1e-1} & \num{9,5} \\ 
			\bottomrule
		\end{tabular} 
	\end{ruledtabular}
\end{table}
\begin{figure}[b]
	\includegraphics[width=0.9\linewidth]{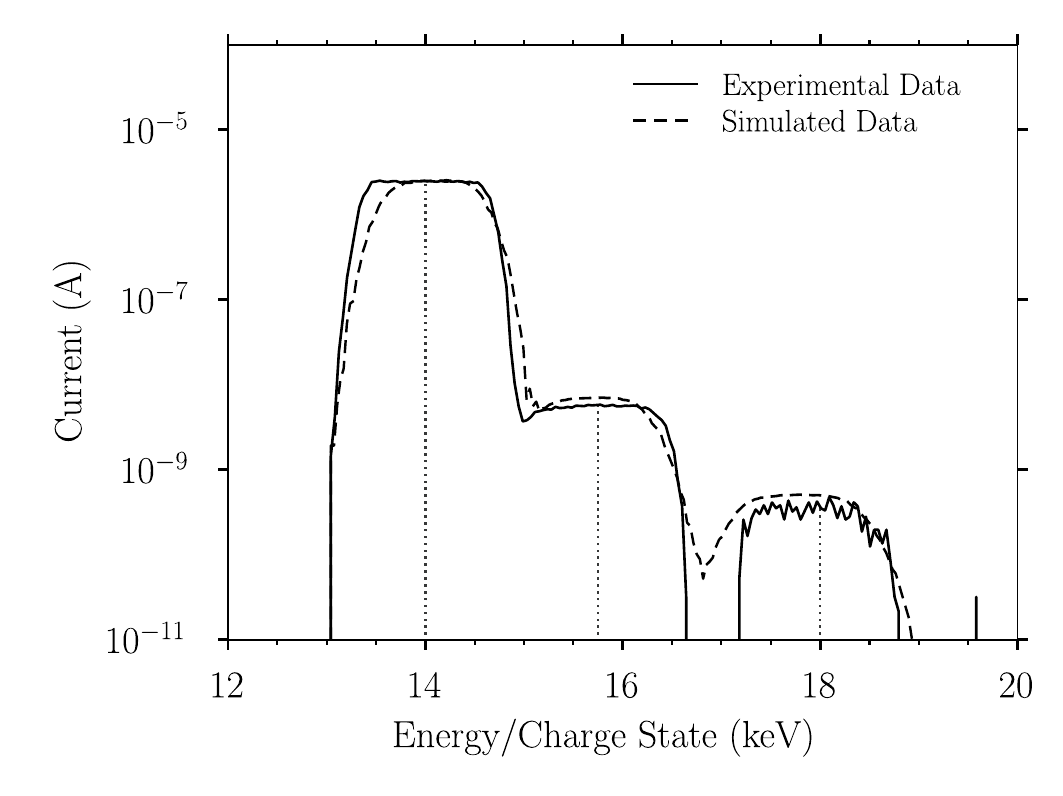}
	\caption{\label{fig:sim_res}Comparison of a measured (solid) and a simulated (dashed) energy scan of an Ar\textsuperscript{9+} ion beam. Vertical dotted lines represent the expected center of each charge state. An energy uncertainty of \SI{0,5}{\percent}, an emittance of \num{30} $\pi$ \si{\milli\metre \, \milli\radian} and a Gaussian shape with \SI{7}{\milli\metre} FWHM were simulated to best reproduce the experimental scan.}
\end{figure}

In \autoref{fig:sim_res} a comparison of the measured scan and a simulated one is shown. Simulated is a \SI{14}{q\kilo\electronvolt} centered starting Gaussian Ar\textsuperscript{9+} ion beam with \SI{0,5}{\percent} energy uncertainty. A transversal emittance of \num{30} $\pi$ \si{\milli\metre \, \milli\radian} and a horizontal and vertical full width at half maximum of \SI{7}{\milli\metre} is chosen, as it reproduced the measured data best. The absolute height of the primary peak as well as the relative heights of the product peaks are scaled according to the values from \autoref{tab:proportion}. Comparison between simulations and measurements, by means of shape and relative peak position, show good agreement. The plateau-like maximum of the scan is narrower with broader edges in the simulation, which results in a higher resolution of about 12, while the peaks are actually broader at their base. The comparison between experiment and simulation shows that the commonly used definition for the resolution has to be treated with care in our case, since as already mentioned the shape of the scans is not Gaussian-like as presumed for this definition of the resolution.

\section{Conclusion}

A simple and compact electrostatic charge state purification system was simulated, built and tested. Its large acceptance allows \SI{100}{\percent} transmission of the primary charge state within the experimental uncertainty for high emittance ($ \leq $ \num{60} $\pi$ \si{\milli\metre \, \milli\radian}) low energy ($ \leq \SI{30}{q\,\kilo\electronvolt} $) ion beams. By utilizing two additional einzel lenses and Matsuda electrodes, we are able to deliver a narrow beam at different focal points. By altering the voltages on the deflector electrodes, an energy scan of the incoming mono-energetic beam can be carried out, effectively resulting in a scan of the charge state. Experiments were performed with different primary charge states at the ARIBE facility at GANIL. The resolution of the purifier was measured to be \num{10,5}. Due to the non-Gaussian peak shape in the energy scans separation is possible for higher charge states than predicted by the measured resolution. The measured results are in good qualitative agreement with our expectations according to numerical beam trajectory calculations. The achieved performance fully satisfies the requirements\citep{Lamour2013} of the FISIC experiments for ion beams up to Ar\textsuperscript{18+}.

\section{Acknowledgements}
We kindly acknowledge funding by ANR under Project-ID ANR-13-IS04-0007 and the LABEX PLAS@PAR project, managed by the ANR, as part of the \textquotedblleft Programme d'Investissements d'Avenir\textquotedblright under the reference ANR-11-IDEX-0004-02. We furthermore want to thank L. Maunoury, V. Toivanen and C. Feierstein for their support during the experiment at the ARIBE facility.

\bibliography{bibliography}

\end{document}